\documentclass[twocolumn,floatfix,showpacs,preprintnumbers,amsmath,amssymb,superscriptaddress,prb]{revtex4}

\usepackage{color}

\usepackage{graphicx}
\usepackage{dcolumn}
\usepackage{bm}

\begin{document}
\title{Low-temperature structural investigations of the frustrated quantum antiferromagnets Cs$_2$Cu(Cl$_{4-x}$Br$_{x}$)}

\author{N. van Well}
\affiliation{Physikalisches Institut, Goethe-Universit\"at Frankfurt, D-60438 Frankfurt am Main, Germany}

\author{K. Foyevtsova}
\affiliation{Quantum Matter Institute, University of British Columbia, Vancouver, British Columbia V6T 1Z4, Canada}
\affiliation{Institut f\"ur Theoretische Physik, Goethe-Universit\"at Frankfurt, D-60438 Frankfurt am Main, Germany}

\author{S. Gottlieb-Sch\"onmeyer}
\affiliation{Physik-Department E21, TU M\"unchen, D-85748 Garching, Germany}

\author{F. Ritter}
\affiliation{Physikalisches Institut, Goethe-Universit\"at Frankfurt, D-60438 Frankfurt am Main, Germany}

\author{R. S. Manna}
\affiliation{Physikalisches Institut, Goethe-Universit\"at Frankfurt, D-60438 Frankfurt am Main, Germany}

\author{B. Wolf}
\affiliation{Physikalisches Institut, Goethe-Universit\"at Frankfurt, D-60438 Frankfurt am Main, Germany}

\author{M. Meven}
\affiliation{Institut f\"ur Kristallographie, RWTH Aachen, D-52056 Aachen, Germany}
\affiliation{J\"ulich Centre for Neutron Science at Heinz Maier-Leibnitz Zentrum, D-85747 Garching, Germany}

\author{C. Pfleiderer}
\affiliation{Physik-Department E21, TU M\"unchen, D-85748 Garching, Germany}

\author{M. Lang}
\affiliation{Physikalisches Institut, Goethe-Universit\"at Frankfurt, D-60438 Frankfurt am Main, Germany}

\author{W. Assmus}
\affiliation{Physikalisches Institut, Goethe-Universit\"at Frankfurt, D-60438 Frankfurt am Main, Germany}

\author{R. Valent\'i}
\affiliation{Institut f\"ur Theoretische Physik, Goethe-Universit\"at Frankfurt, D-60438 Frankfurt am Main, Germany}

\author{C. Krellner}
\email{krellner@physik.uni-frankfurt.de}
\affiliation{Physikalisches Institut, Goethe-Universit\"at Frankfurt, D-60438 Frankfurt am Main, Germany}
\date{\today}

\begin{abstract}
Powder X-ray diffraction (PXRD) and single-crystal neutron scattering were used to study in detail the structural properties of the Cs$_2$Cu(Cl$_{4-x}$Br$_{x}$) series, good realizations of layered triangular antiferromagnets. Detailed temperature-dependent PXRD reveal a pronounced anisotropy of the thermal expansion for the three different crystal directions of the orthorhombic structure without any structural phase transition down to 20\,K. Remarkably, the anisotropy of the thermal expansion varies for different $x$, leading to distinct changes of the geometry of the local Cu environment as a function of temperature and composition. The refinement of the atomic positions confirms that for $x=1$ and 2, the Br atoms  occupy distinct halogen sites in the [CuX$_4$]-tetrahedra (X = Cl, Br). 
The precise structure data are used to calculate the magnetic exchange couplings using density functional methods for $x=0$. We observe a pronounced temperature dependence of the calculated magnetic exchange couplings, reflected in the strong sensitivity of the magnetic exchange couplings on structural details. These calculations  are in good agreement with the experimentally established values for Cs$_2$CuCl$_4$ if one takes the low-temperature structure data as a starting point.  

\end{abstract}

\maketitle

\section{\label{sec:Introduction} Introduction}
Cs$_2$CuCl$_{4}$ and Cs$_2$CuBr$_{4}$ are experimental realizations  of two-dimensional triangular-lattice spin-$\frac{1}{2}$ Heisenberg antiferromagnets  which have been intensively studied in recent years due to their unconventional magnetic properties at low temperatures. Despite their structural similarities, Cs$_2$CuCl$_{4}$ shows a pronounced spin-liquid behavior at low magnetic fields as well as a Bose-Einstein condensation of magnons at a critical field,\cite{Coldea:2001, Coldea:2002, Coldea:2003, Radu:2005a} whereas Cs$_2$CuBr$_{4}$ has localized magnetic excitations with well-defined plateaus in the magnetization. \cite{Ono:2003, Ono:2005} This situation indicates that the magnetic properties are highly sensitive to small details of the crystal structure, which, so far, has been  established  only at room temperature. \cite{Bailleul:1991, Morosin:1960} 

The important magnetic units in these systems are Jahn-Teller distorted (CuX$_4$)-tetrahedra (X = Cl, Br) which are arranged in layers separated by the alkali atoms. A deep understanding of the  relationship between crystal structure and electronic and magnetic behavior in the series Cs$_2$Cu(Cl$_{4-x}$Br$_{x}$) requires a detailed  structural analysis as a function of $x$ and temperature. Therefore, we started the crystal growth of the mixed system Cs$_2$Cu(Cl$_{4-x}$Br$_{x}$) with a systematic substitution of Cl by Br for $0 \leq  x \leq  4$. It was shown that the orthorhombic structure type ($Pnma$) characterized by  triangular-lattice spin-$\frac{1}{2}$-layers can be obtained from aqueous solution over the whole range of $x$ if the growth temperature is above $50^{\circ}$C. \cite{Kruger:2010} The magnetic properties and sterical considerations suggested distinct magnetic regimes through a site-selective substitution of the Br atoms on the three different crystallographic halogen positions. \cite{Kruger:2010, Cong:2011}   However, up to now no direct structural evidence for such a scenario has been given and the detailed crystal structures at low temperature have not been  available except the lattice parameters for $x=0$. \cite{Coldea:2001, Coldea:2002}

The purpose of the present work is to gain detailed information on  the structural properties of  Cs$_2$Cu(Cl$_{4-x}$Br$_{x}$) at low temperatures by means of X-ray and neutron diffraction experiments.  In particular, we are specially interested in identifying any structural phase transition that might occur upon lowering the temperature as the orthorhombic structure was found to be metastable for $0\leq x<4$.\cite{Kruger:2010} Furthermore, we determine accurately the crystal structure parameters for $x=0$, $1$, $2$, and $4$. In addition, we perform density functional theory (DFT) calculations using the identified crystal structures and investigate the various superexchange couplings between Cu spins.  In previous DFT calculations it was demonstrated that  the superexchange couplings sensitively depend on fine structural details of the local Cu environment given by  the [CuX$_4$]-tetrahedra (X denotes the halogen atom, Cl or Br). \cite{Foyevtsova:2011} This indicates that even in the absence of a structural phase transition, magnetic interactions in  Cs$_2$Cu(Cl$_{4-x}$Br$_{x}$) might be strongly modified by temperature-induced structural variations. One of the main findings of the present study is that the latter effect is indeed strong in Cs$_2$CuCl$_{4}$, and the DFT calculations with the correct low-temperature crystal structure can reproduce the experimental coupling constants. 

The paper is organized in the following way: In Sec.~\ref{sec:Exp} we provide experimental details of the crystal growth and the structural characterization. The low-temperature PRXD data of the Cs$_2$Cu(Cl$_{4-x}$Br$_{x}$) mixed system are presented in Sec.~\ref{sec:TTXRD}, followed by a detailed structural analysis of the local Cu environment in Sec.~\ref{sec:tetrahedra}. Finally, we discuss the temperature dependence of the spin superexchange coupling constants of Cs$_2$CuCl$_{4}$ in Sec.~\ref{sec:DFT} by performing DFT calculations using the 300 and 20 K crystallographic data, followed by our conclusions.

\begin{figure}
    \includegraphics[width=0.9\columnwidth]{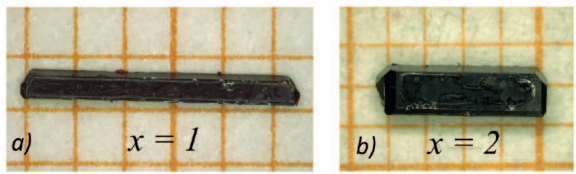}
\caption{\label{fig1} (Color online) Optical images of the as-grown orthorhombic ($Pnma$) crystals, a) Cs$_2$CuCl$_3$Br$_1$ and b) Cs$_2$CuCl$_2$Br$_2$, obtained from aqueous solution at 50$^{\circ}$C. The given scale is 1 mm.}%
\end{figure}

\section{\label{sec:Exp} {Experimental Details}}
The Cs$_2$Cu(Cl$_{4-x}$Br$_{x}$) crystals with an orthorhombic structure were grown from aqueous solution at $50^{\circ}$C with the evaporation method. \cite{Kruger:2010} Typical crystals for $x=1$ and $x=2$ are shown in Fig. \ref{fig1}, their growth time lasts around two months. 

\begin{figure}
     \includegraphics[width=\columnwidth]{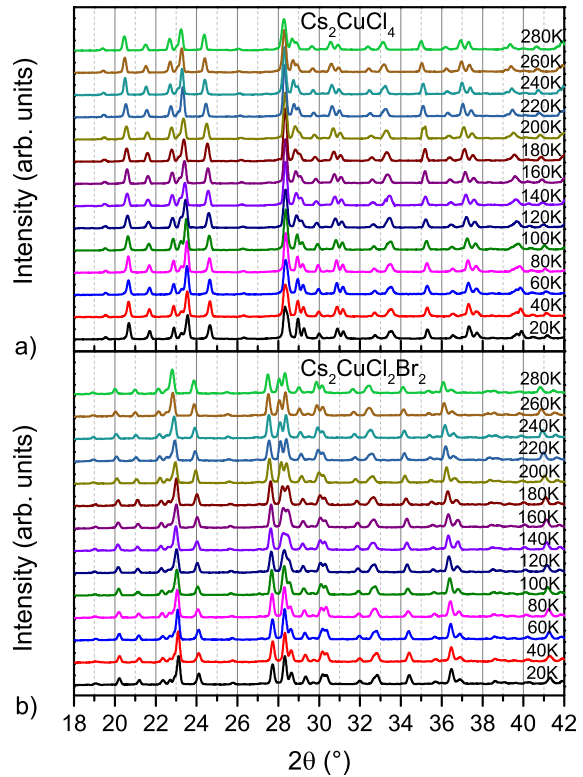}
\caption{\label{fig2} (Color online) PXRD data from 280 to 20\,K exemplarily shown for a) Cs$_2$CuCl$_{4}$ and b) Cs$_2$CuCl$_2$Br$_2$. No structural phase transition for any of the materials was observed between 280 and 20 K. The high quality of these data allows for the determination of the thermal expansion coefficients.}%
\end{figure}

For the low-temperature powder X-ray diffraction (PXRD) a Siemens D500 diffractometer with Cu$K_{\alpha}$-radiation was used. Cooling of the powder was realized with a closed-cycle helium refrigerator. The measurements were started at 280 K followed by subsequent cooling in 20 K steps down to 20 K (see Fig.~\ref{fig2}). At each temperature, $T$, the sample holder with the powder was thermalized, prior to the measurement scan, employing an in-situ temperature controller. Each powder sample was obtained by milling one single crystal and mixing it with powder of silicon. The latter was used as an internal standard for the temperature dependent correction of the zero shift and the displacement of the powder. Structural parameters were refined from the PXRD data using the GSAS Suite of Rietveld programs. \cite{GSAS} The determination of the low-temperature structure was done for four different concentrations with nominal values of $x = 0, 1, 2, 4$. Microprobe analysis confirmed that these nominal values are close to the actual ones (see Ref.~\onlinecite{Kruger:2010}). 

The measurements of the uniaxial thermal expansion coefficient $\alpha = l^{-1}(\partial l / \partial T)$, where $l$ denotes the sample length, were conducted by employing a high-resolution capacitive dilatometer with a maximum resolution of $\Delta l/l = 10^{-10}$, built after Ref. \onlinecite{Pott:1983}.

Neutron scattering experiments at 300 K were carried out for one single crystal with $x=1$ at the 4-circle-diffractometer HEIDI at the FRM II. A full sphere up to $2\Theta= 65^{\circ}$ was measured  with a neutron wavelength of $\lambda=0.87$\,\AA\, (in total 1940 reflexes). The neutron data was refined using the software Jana2006. \cite{Jana2006}

\section{\label{sec:TTXRD} {Low-temperature structural characterization}}

PXRD was measured and analyzed from 280 to 20 K for selected members of the Cs$_2$Cu(Cl$_{4-x}$Br$_{x}$) series. Besides the end members ($x = 0$ and 4), the compositions Cs$_2$CuCl$_3$Br$_1$ and Cs$_2$CuCl$_2$Br$_2$ were chosen for a detailed structural characterization, because for these compounds a partially ordered distribution of the halogen components with respect to the different crystallographic positions was suggested. \cite{Cong:2011} The complete PXRD data for two of the four different concentrations are shown in Fig.~\ref{fig2}. These data clearly indicate that the orthorhombic structure ($Pnma$) of the crystals remains stable down to 20 K without any indication for a structural phase transition between 280 and 20 K. The high quality of these data allows for a detailed analysis of the structural parameters as function of temperature.

\begin{figure}
     \includegraphics[width=0.75\columnwidth]{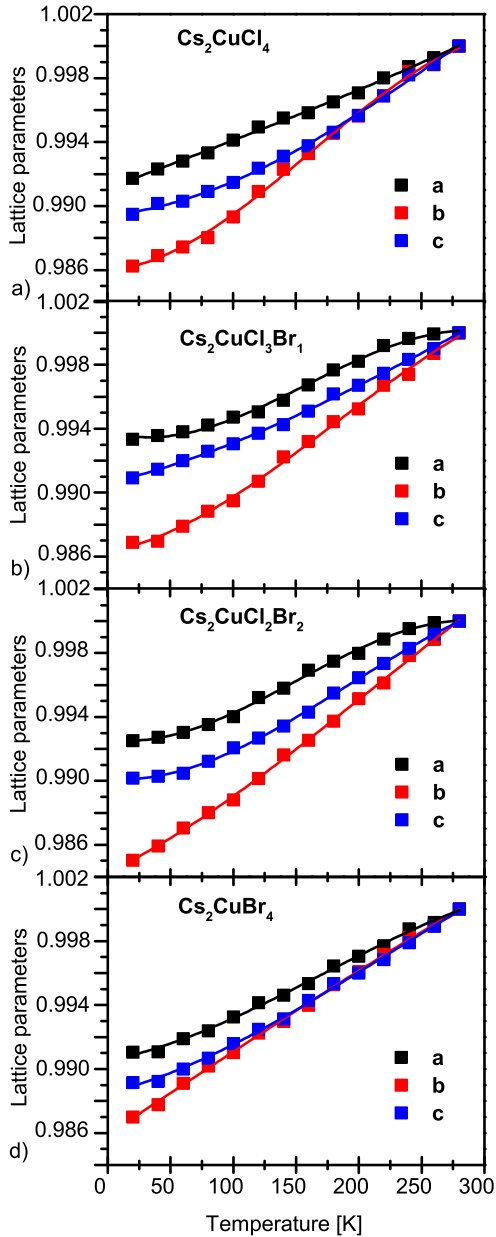}
\caption{\label{fig3} (Color online) Normalized lattice parameters for a) Cs$_2$CuCl$_{4}$, b) Cs$_2$CuCl$_3$Br$_1$, c) Cs$_2$CuCl$_2$Br$_2$, and d) Cs$_2$CuBr$_{4}$, plotted as $l_i(T)/l_i(280\,K)$. The size of each point exceeds the statistical error. The $T$-dependence was fitted with a third-order polynomial function (solid lines), which were used to calculate the coefficients of the thermal expansion.}%
\end{figure}

\begin{figure}
     \includegraphics[width=0.75\columnwidth]{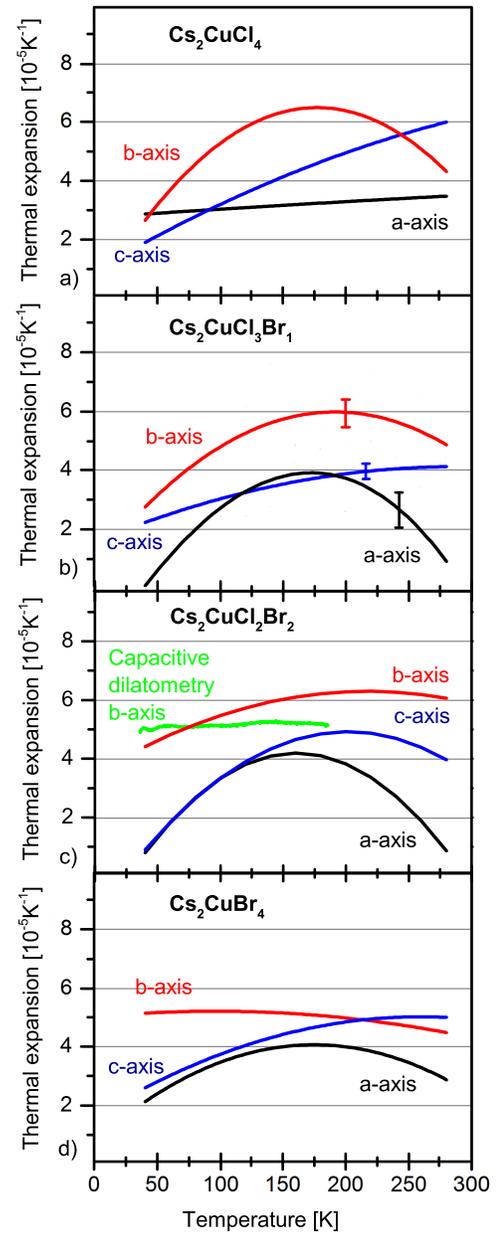}
\caption{\label{fig4} (Color online) $T$-dependence of the linear thermal expansion coefficients for a) Cs$_2$CuCl$_{4}$, b) Cs$_2$CuCl$_3$Br$_1$, c) Cs$_2$CuCl$_2$Br$_2$, and d) Cs$_2$CuBr$_{4}$. Typical errors correspond to the error bars shown in b). The thermal expansion coefficients show a pronounced anisotropy for the three different crystal directions $a$, $b$, $c$ and vary considerably with $x$.  For $x=2$, we compare the data determined from the lattice parameters with dilatometry data.}%
\end{figure}

\begingroup
\squeezetable
\begin{ruledtabular}
\begin{table*}
\begin{center}
\caption{\label{tab1} Lattice parameters and atomic positions for $x=0,1,2,4$ at 20 K, refined from PXRD data.}
\begin{tabular}{|l|c|c|c|c|c|c|c|c|c|c|c|c|}
\hline 
&  \multicolumn{3}{c}{Cs$_2$CuCl$_{4}$}   \vline & \multicolumn{3}{c}{Cs$_2$CuCl$_3$Br$_1$}  \vline &\multicolumn{3}{c}{Cs$_2$CuCl$_2$Br$_2$}   \vline & \multicolumn{3}{c}{Cs$_2$CuBr$_{4}$} \vline \\ 
\hline  
$a$ (\AA) &  \multicolumn{3}{c}{9.675(3)}   \vline & \multicolumn{3}{c}{9.819(7)}  \vline &\multicolumn{3}{c}{9.948(5)}   \vline & \multicolumn{3}{c}{10.069(7)} \vline \\ 
\hline
$b$ (\AA) &  \multicolumn{3}{c}{7.496(2)}   \vline & \multicolumn{3}{c}{7.552(5)}  \vline &\multicolumn{3}{c}{7.633(3)}   \vline & \multicolumn{3}{c}{7.839(5)} \vline \\ 
\hline
$c$ (\AA) &  \multicolumn{3}{c}{12.264(3)}   \vline & \multicolumn{3}{c}{12.408(1)}  \vline &\multicolumn{3}{c}{12.560(2)}   \vline & \multicolumn{3}{c}{12.758(1)} \vline \\ 
\hline  \hline
Atom	& $x$ & $y$ & $z$ & $x$ & $y$ & $z$ & $x$ & $y$ & $z$ & $x$ & $y$ & $z$ \\  \hline 
Cs $4c$	& 0.132(3) & 0.25 &0.102(3) &0.131(1) & 0.25 & 0.101(1) & 0.128(1) & 0.25 & 0.097(1) & 0.124(1) &0.25 & 0.106(1) \\ 
Cs $4c$	& 0.502(3) & 0.25 &0.824(3) &0.510(1) & 0.25 & 0.816(1)& 0.510(1) & 0.25 & 0.826(1) & 0.496(1) &0.25 & 0.831(1) \\ \hline
Cu $4c$	& 0.221(8)& 0.25 &0.410(7) &0.234(4) & 0.25 & 0.410(3)& 0.243(3) & 0.25 & 0.421(3) & 0.224(4) &0.25 & 0.411(3) \\ \hline
Cl1 $4c$	& 0.332(1)& 0.25 &0.575(1) &0.317(4) & 0.25 & 0.589(4)& - & - & - &- &- &- \\ 
Cl2 $4c$	& -0.008(1)& 0.25 &0.381(1) &- &- &- &- &- &- &- &- &- \\ 
Cl3 $8d$	& 0.298(1)& -0.032(1)&0.357(1) &0.288(6) &0.006(6) &0.357(3) & 0.291(3) & -0.001(4) &0.352(3)  &- &- &- \\ \hline
Br1 $4c$	&- &- &- &- &- &- & 0.345(2) & 0.25 & 0.580(2) &0.343(2) &0.25 & 0.585(2)\\ 
Br2 $4c$	&- &- &- & -0.008(2) & 0.25 &0.384(2) & -0.011(2) & 0.25 &0.374(1)  &-0.006(2) &0.25 &0.377(2) \\ 
Br3 $8d$	&- &- &- &- &- &- &- &- &- &0.298(1) &-0.013(3) &0.357(1) \\ 
	 \hline
\hline
$\chi^2_{red}$ &  \multicolumn{3}{c}{3.20}   \vline & \multicolumn{3}{c}{3.12}  \vline &\multicolumn{3}{c}{2.95}   \vline & \multicolumn{3}{c}{2.91} \vline \\
wRp &  \multicolumn{3}{c}{0.095}   \vline & \multicolumn{3}{c}{0.098}  \vline &\multicolumn{3}{c}{0.087}   \vline & \multicolumn{3}{c}{0.085} \vline \\
Rp&  \multicolumn{3}{c}{0.072}   \vline & \multicolumn{3}{c}{0.069}  \vline &\multicolumn{3}{c}{0.062}   \vline & \multicolumn{3}{c}{0.065} \vline \\
\hline  
\end{tabular}
\end{center}
\end{table*}
\end{ruledtabular}
\endgroup

From the temperature-dependent PXRD data in Fig.~\ref{fig2} we derived the lattice parameters $l_i = a,b,c$. The variation of these parameters with temperature is shown in Fig.~\ref{fig3} by plotting the $l_i(T)$ data normalized to their values at 280 K, $l_i(T)/l_i(280\,\textnormal{K})$. The $T$-dependence of the lattice parameters was fitted with a third-order polynomial function, because inflection points are apparent. These polynomial functions are later used to determine the coefficients of the thermal expansion and are shown as solid lines in Fig.~\ref{fig3}. It is obvious from this figure that $l_i(T)/l(280\,\textnormal{K})$ is anisotropic and changes drastically with increasing $x$. For $x=0$, $a(T)$ is linear in temperature down to 20 K, which markedly changes for $x =1$ and 2. For $b(T)$ this trend is reversed, since for $x=0$ we observe a strong deviation from a linear-$T$ dependence while as $x$ increases this dependence becomes more linear. The overall anisotropy of the normalized lattice parameters at 20 K is small for $x=4$, but larger for $x=2$, due to a larger $a(20\,\textnormal{K})/a(280\,\textnormal{K})$ and a smaller $b(20\,\textnormal{K})/b(280\,\textnormal{K})$ value. This is a further indication for a selective occupation of the halogen positions by Cl and Br as will be discussed in more detail in Sec.~\ref{sec:tetrahedra}. Another reason for the pronounced anisotropy of the $T$-dependence of the lattice parameters could be related to the fact that for $0\leq x<4$ the orthorhombic crystal structure is metastable, meaning that a structurally different polymorph exists, which is energetically more stable. This was explicitly shown for $1\leq x\leq 2$, where a tetragonal crystal structure ($I4/mmm$) forms, when the crystals are grown from aqueous solution at lower temperatures. \cite{Kruger:2010} Also for $x=0$, we have found a tetragonal variant of Cs$_2$CuCl$_{4}$, if the crystal growth is carried out at $8^{\circ}$C. \cite{Kruger:2010, vanWell:2014} For $x=4$, the smaller anisotropy of $l_i(T)/l_i(280\,K)$ indicates a more stable crystal structure, and indeed no different polymorph is known to exist so far for Cs$_2$CuBr$_{4}$.

The coefficients of the linear thermal expansion, $\alpha_i$,  are determined by taking the derivative of the normalized lattice parameters with respect to $T$, i.e., 
\begin{equation*}
\alpha_i = \frac{1}{l_i(280\,K)}\bigl{(}\frac{\partial l_i(T)}{\partial T}\bigr{)}_p \,.
\end{equation*}
We approximated these coefficients by using the fitted third-order polynomial function for $l_i(T)/l_i(280\,\textnormal{K})$ in Fig.~\ref{fig3}. As a consequence the accuracy at the high- and low-temperature ends (20 and 280\,K) is limited. The $T$-dependence of these calculated coefficients are shown in Fig.~\ref{fig4} for the four different concentrations. Representative error bars are shown in Fig.~\ref{fig4}b, which are estimated from the deviation of the polynomial function from the data points in Fig.~\ref{fig3}. In Fig.~\ref{fig4} the above discussed anisotropy is even more visible. For $x=0$, a pronounced maximum at around 170\,K is apparent for the $b$-axis (Fig.~\ref{fig4}a), whereas the coefficient for the $a$-axis is nearly constant. Going to $x=1$ (Fig.~\ref{fig4}b), the largest change appears for the $a$-direction, where a broad maximum is observed above 150\,K. We attribute this strongly anisotropic change to a preferred occupation of the Br-atoms on the X2-site, which is the crystallographic site pointing along the $a$-direction of the tetrahedron (cf. Fig.~\ref{fig5}), because statistically distributed Br over the three different lattice sites would probably lead to similar changes for the three thermal expansion coefficients. Fig.~\ref{fig4}c presents the data for $x=2$, here the most obvious change compared to the $x=1$ sample appears along the $c$-direction. Also this behavior is in agreement with the site-selective occupation, because for $1\leq x\leq 2$, the X1-site is occupied by the Br-atoms, which is the crystallographic site along the $c$-direction of the tetrahedron (cf. Fig.~\ref{fig5}). Finally, in Fig.~\ref{fig4}d the coefficients of the thermal expansion are shown for Cs$_2$CuBr$_{4}$ with an overall reduced anisotropy.

For Cs$_2$CuCl$_{4}$, we can compare our data with thermal expansion coefficients, reported in the literature (Fig.~3 in Ref.~\onlinecite{Tylczynski:1992}). These dilatometry data (not shown) were measured using a push-rod method \cite{Tylczynski:1992} and the agreement with our data is satisfactorily good. Particularly, a broad maximum for the $b$-direction ($\alpha_{33}$ in the notation of Ref.~\onlinecite{Tylczynski:1992}) was observed around 180\,K in accordance with our results in Fig.~\ref{fig4}a. In addition, we determined the thermal expansion coefficent for Cs$_2$CuCl$_2$Br$_2$ along the $b$-axis using capacitive dilatometry. Within the errors, this data is in agreement with the results from PXRD as shown in Fig.~\ref{fig4}c. 

\begin{ruledtabular}
\begin{table}
\begin{center}
\caption{\label{tab1n} Atomic positions for Cs$_2$CuCl$_3$Br$_1$ at 300 K, refined from single-crystal neutron diffraction and PXRD data.}
\begin{tabular}{|l|c|c|c|}
\hline
&  \multicolumn{3}{c}{Neutron data} \vline \\ 
\hline  
Atom	& $x$ & $y$ & $z$ \\  \hline 
Cs $4c$	& 0.13132(16) & 0.25 &0.10183(13)\\ 
Cs $4c$	& 0.49383(15)& 0.25 &0.82403(11)\\ \hline
Cu $4c$	& 0.23164(9)& 0.25 &0.41767(7)\\ \hline
Cl1 $4c$	& 0.34222(9)& 0.25 &0.57575(8)\\ 
Br2 $4c$	& 0.00292(10)& 0.25 &0.38095(8)\\ 
Cl3 $8d$	& 0.29338(7)& 0.01202(7)&0.35627(6)\\ \hline
GOF &  \multicolumn{3}{c}{1.29}\vline \\ \hline \hline
&  \multicolumn{3}{c}{PXRD data} \vline \\ \hline
Atom	& $x$ & $y$ & $z$ \\  \hline 
Cs $4c$	&  0.132(1) & 0.25 &0.098(1)\\ 
Cs $4c$	&  0.510(1)& 0.25 &0.826(1)\\ \hline
Cu $4c$	& 0.221(3)& 0.25 &0.425(3)\\ \hline
Cl1 $4c$	& 0.337(4)& 0.25 &0.562(4) \\ 
Br2 $4c$	& -0.005(2)& 0.25 &0.377(1) \\ 
Cl3 $8d$	& 0.291(3)& 0.003(5)&0.367(3)\\ \hline
GOF &\multicolumn{3}{c}{3.54} \vline \\
\hline  
\end{tabular}
\end{center}
\end{table}
\end{ruledtabular}

\begin{ruledtabular}
\begin{table*}
\begin{center}
\caption{\label{tab3}Geometrical characteristics of the [CuX$_4$]-tetrahedra (X = Cl, Br) at 20\,K, the given parameters are depicted in Fig.~\ref{fig5}c.}
\begin{tabular}{|l|c|c|c|c|}
\hline 
& Cs$_2$CuCl$_{4}$ & Cs$_2$CuCl$_3$Br$_1$ & Cs$_2$CuCl$_2$Br$_2$ & Cs$_2$CuBr$_{4}$ \\ 
\hline  \hline 
$\alpha$(Cu-X3-X3) $(^{\circ})$, 280\,K &151.9 & 152.9 &159.3 & 155.8 \\  \hline 
$\alpha$(Cu-X3-X3) $(^{\circ})$, 20\,K  & 154.9 & 155.1 & 152.8 & 153.9  \\  \hline \hline 
d(Cu-X3) (\AA), 280\,K  & 2.204(8) & 2.121(4) & 2.312(3) & 2.307(4) \\  \hline 
d(Cu-X3) (\AA), 20\,K  & 2.336(8) & 2.032(4) & 2.154(3) & 2.300(4) \\  \hline \hline 
d(X3-X3) (\AA), 280\,K  & 3.721(1) & 3.871(6) & 3.423(3) & 3.746(4) \\  \hline 
d(X3-X3) (\AA), 20\,K  & 3.262(1) & 3.865(6) & 3.799(3) & 3.721(4) \\  \hline \hline 
d(Cu-X3-X3-Cu) (\AA), 280\,K &8.129(11) & 8.113(8) &8.047(5) & 8.360(6) \\  \hline 
d(Cu-X3-X3-Cu) (\AA), 20\,K  &7.934(11) & 7.929(8) &8.107(5) & 8.321(6)  \\  \hline 
\end{tabular}
\end{center}
\end{table*}
\end{ruledtabular}

\section{\label{sec:tetrahedra} {Analysis of the [C\MakeLowercase{u}X$_4$]-tetrahedra}}
\begin{figure}
   \mbox{
    \includegraphics[width=\columnwidth]{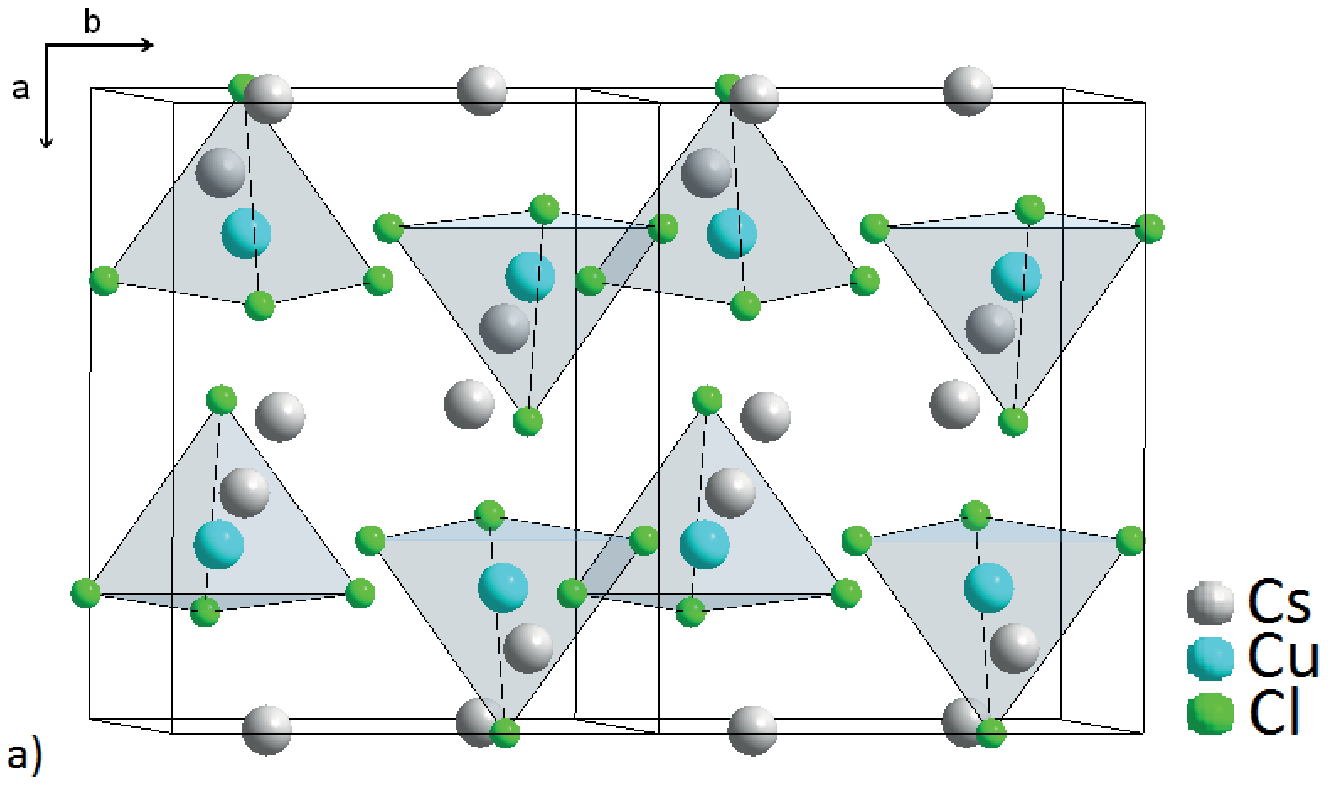}
   }
   \mbox{
    \includegraphics[width=0.75\columnwidth]{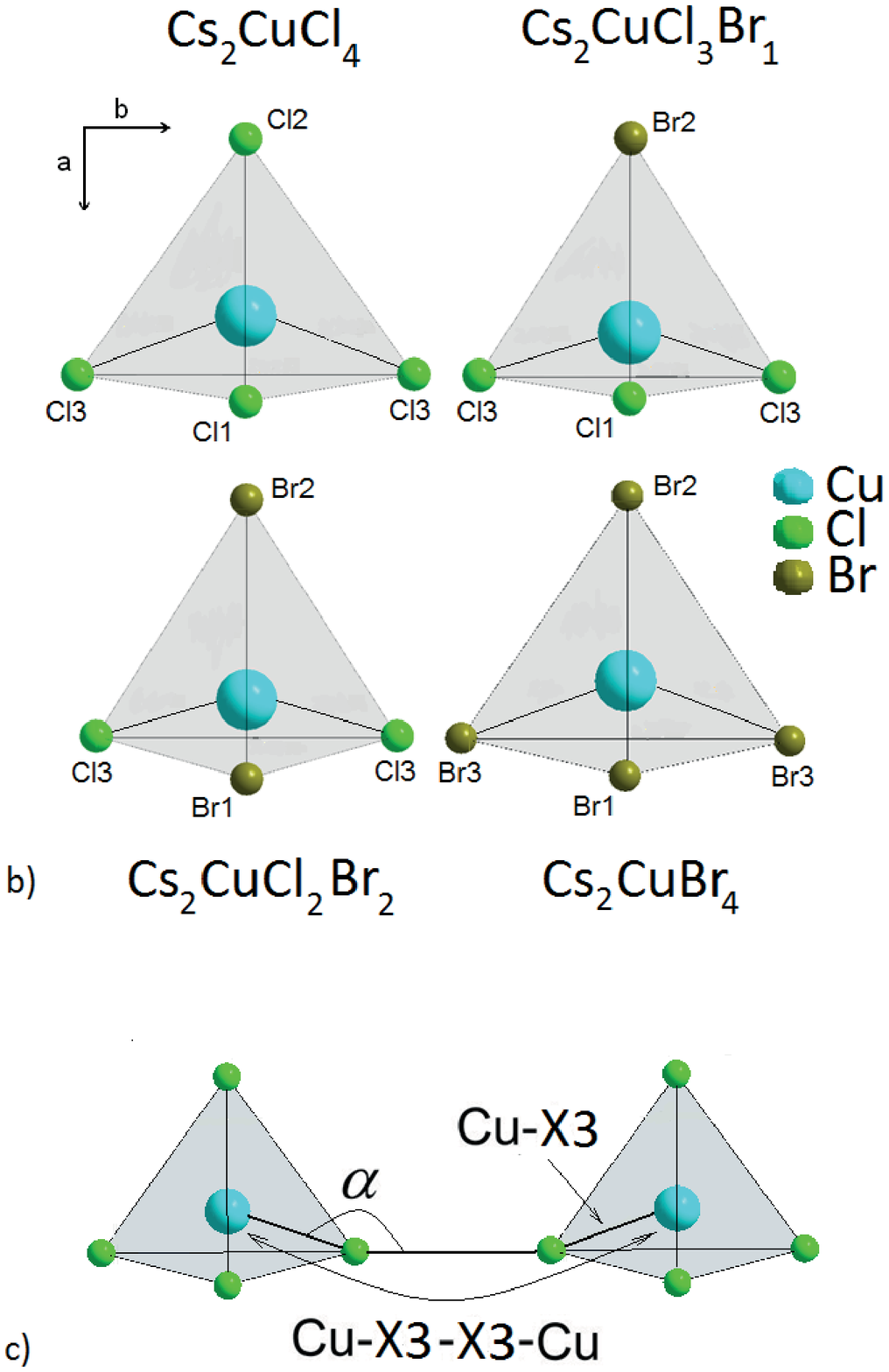}
   }
\caption{\label{fig5} (Color online) a) Crystal structure for $x=0$ with the [CuX$_4$]-tetrahedra for two neighboring unit cells along the $b$-direction. b) [CuX$_4$]-tetrahedron for the four investigated compositions calculated from the PXRD data at 20\,K. All tetrahedra are shown in the $ab$ plane and are normalized to their height.  b) Designation of the  magnetic exchange path between two consecutive [CuX$_4$]-tetrahedra along the $b$-direction with the angle $\alpha$ as well as the Cu-X3 and Cu-X3-X3-Cu distances used in Table~\ref{tab3}.}%
\end{figure}

The structurally most important unit for the magnetic exchange interactions in these materials is the [CuX$_4$]-tetrahedron (X = Cl, Br) which determines the local environment of Cu (see Fig.~\ref{fig5}a). In this section we will analyze the form of the tetrahedron in more detail for the different compositions of Cs$_2$Cu(Cl$_{4-x}$Br$_{x}$) with $x = 0, 1, 2, 4$, refined from the low-temperature PXRD data. One important  finding of our structural analysis is the site-selective occupation of the Br-atoms on the three different halogen sites, illustrated in Fig.~\ref{fig5}b. This is already apparent from the development of the lattice parameters with $x$, shown in the first lines of Table~\ref{tab1}. Going from $x=0$ to 1, the largest relative change occurs along the $a$-direction, whereas the relative change of $b$ is much smaller. Going from $x=1$ to 2, the smallest increase is again for the $b$-direction, but much larger for $a$ and $c$. The opposite is observed from $x=2$ to 4, here the largest increase is found for the $b$-direction, but much smaller relative changes of the $a$ and $c$-direction. Looking at the corners of the tetrahedron and its location within the crystal structure, it is clear that the X1-site will have the strongest influence for the $c$-parameter (front corner), the X2-site for  $a$ (top corner), and the X3-sites for $b$ (side corners) of the orthorhombic crystal structure as depicted in Fig.~\ref{fig5}b. Similar behavior was observed in the room-temperature PXRD data in Ref.~\onlinecite{Kruger:2010}.

\begin{figure}
\includegraphics[width=\columnwidth]{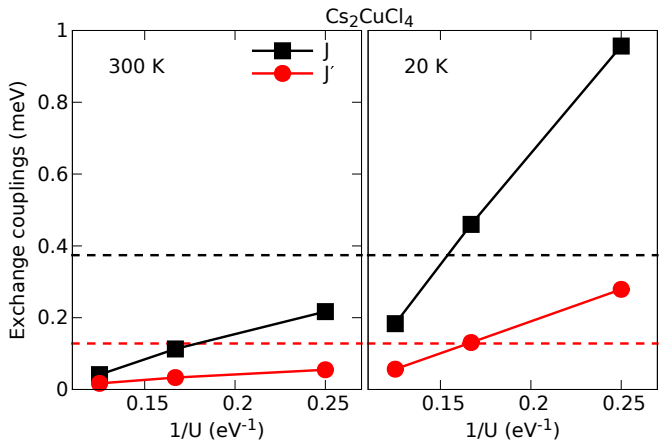}%
\caption{\label{fig6} (Color online) Spin exchange couplings for Cs$_2$CuCl$_{4}$ determined with density functional methods using two different sets of structural data, obtained at 300 and  20\,K. Dashed lines denote the experimental findings for $J$ and $J'$. \cite{Coldea:2001}}%
\end{figure}

The atomic positions  refined from the PXRD data at 20 K  are summarized in Table~\ref{tab1}. In a first attempt, a random distribution of the Cl and Br atoms was supposed for $x=1$ and 2. However, with this configuration it was not possible to get a reasonable agreement between the measured and  simulated data. Therefore, in a second step, an ordered model with Br atoms on the X2 site for the Cs$_2$CuCl$_3$Br$_1$ compound and with the Br atoms on the X1 and X2 site for Cs$_2$CuCl$_2$Br$_2$ was assumed, as indicated from the development of the lattice parameters discussed above. With such a configuration, good agreement could be achieved with much lower $\chi_{\textnormal{red}}$ values. Furthermore, no peculiarities were observed in the isotropic displacement parameters for the different atomic positions. Therefore, in Table~\ref{tab1} we have split the halogen positions for Cl and Br. However, within the resolution of our PXRD data we cannot completely rule out some Cl-Br site-disorder in these compounds. From several measurements on different powdered crystals we estimate an upper boundary of this site-disorder of about 10\%. This means for Cs$_2$CuCl$_2$Br$_2$ that at most every $10^{th}$ Br-atom might sit on a X3 site. We like to note that for crystals grown directly from the melt at around $450^{\circ}$C (Bridgman-method), we could not observe this site-selective occupation of the Br-atoms, but only a random distribution of the halogen atoms on the three different crystallographic positions. \cite{vanWell:2014} To confirm our extended temperature-dependent PXRD measurements, we carried out neutron diffraction on a single crystal with dimensions of $2\times 2 \times 1$\,mm$^3$ of Cs$_2$CuCl$_3$Br$_1$ (cf. Table~\ref{tab1n}). Obviously, the positions of the atoms could be refined with higher resolution compared to the PXRD data as shown in Table~\ref{tab1n}, where we compare the structural data at 300\,K for both methods. More important is the fact that within the given errors both methods reveal similar crystallographic positions. Furthermore, the selective occupation of Br on the X2 site, was also observed in the neutron data.

In Fig.~\ref{fig5}b we present the [CuX$_4$]-tetrahedra for the four different Br-concentrations determined from the PXRD data at 20\,K. The tetrahedra for $x=0$ and 4 are distorted due to the Jahn-Teller effect, which is well established in these materials. On top of the Jahn-Teller distortion, the site-selective occupation of the Br-atoms leads to an additional deformation for $x= 1$ and 2 evident from  Fig.~\ref{fig5}b. Cs$_2$CuCl$_3$Br$_1$ has the Br atom on the crystallographic X2 site, leading to a [CuCl$_3$Br$_1$]-tetrahedron  elongated along $a$, but compressed in its $b$ direction. Cs$_2$CuCl$_2$Br$_2$ with the Br-atoms on the X1 and X2 sites  presents a [CuCl$_2$Br$_2$]-tetrahedron  stretched along $a$ and compressed along $b$. This non-linear change with $x$ of the tetrahedra has direct consequences on the low-temperature magnetic properties of these materials, which are determined mainly by the two in-plane exchange interactions $J$ and $J'$. A first estimate of the dominant $J$ can be gained by considering the length of the superexchange path along the $b$ direction, depicted in Fig.~\ref{fig5}c. This Cu-X3-X3-Cu distance is determined by the bonding angle $\alpha$ and the distances Cu-X3 as well as X3-X3. These parameters are summarized in Table~\ref{tab3}, where the non-linear development with $x$ becomes apparent, in combination with a strongly varying temperature dependence (including positive and negative $T$-coefficients). In the last line of Table~\ref{tab3} the length of the spin superexchange path from Cu to Cu along the $b$-direction at 20 K is shown.  From this data one would expect a nearly unchanged $J$ value for $x=0$ and 1, followed by an increase of $J$ towards $x=4$. This structural argument is in agreement with susceptibility data on the same type of single crystals. \cite{Cong:2011} There it was shown that the maximum at around 3\,K, which is primarily determined through $J$, is nearly unchanged for $0\leq x\leq 1.5$, followed by a linear increase for $x\geq 2$. 

In order to study the influence of these strong structural variations with $x$ and $T$ on the magnetic interactions in more detail, a more accurate calculation of the spin superexchange coupling constants using density functional methods would be insightful. In the next section, we present DFT calculations for Cs$_2$CuCl$_{4}$ with the low-$T$ structural data as input parameters  and compare them with our earlier results, which were obtained from the room-temperature structure data. 

\section{\label{sec:DFT} {DFT-based derivation of exchange-coupling constants}}
To derive the spin superexchange coupling constants in Cs$_2$CuCl$_{4}$ within DFT, we employ the full-potential linearized augmented plane waves method as implemented in the WIEN2k code. \cite{FPLO} The exchange and correlation effects are treated within the generalized gradient approximation \cite{GGA} with an additional orbital-dependent term that mimics the on-site Coulomb repulsion between the Cu 3$d$ electrons. \cite{GGA+U} The coupling constants are calculated from the total energies of different Cu-spin configurations, assuming Ising-like interactions, see Foyevtsova \textit{et al.} (Ref.~\onlinecite{Foyevtsova:2011}) for more technical details.

In Fig.~\ref{fig6}, we compare the spin superexchange couplings in Cs$_2$CuCl$_{4}$ calculated using the experimental crystal structures at 300 and 20 K from the presented PXRD data. No further structural relaxation was performed within DFT. For both temperatures, the leading magnetic interactions are by far the nearest-neighbor couplings $J$ and $J'$ within the triangular lattice. We calculated these two superexchange paths for three values of the on-site Coulomb repulsion $U=4, 6$ and 8 eV. The realistic $U$ value is close to 6 eV ($1/U = 0.17$ 1/eV). The temperature dependence of $J$ and $J'$, however, is remarkable, because these couplings from the 20 K structure are by a factor of $\sim 3$ larger than those from the 300 K structure, while their ratio $J'/J\sim 0.3$ remains roughly unchanged. Noticeably, the exchange couplings calculated from the 20 K structure are in good agreement with experimental estimates (based on neutron scattering data at 0.3 K from Ref.~\onlinecite{Coldea:2001}), marked in Fig.~\ref{fig6} with horizontal lines of respective colors. 

Comparing the structural data at 300 and 20\,K, we can identify structural changes that can be responsible for such a pronounced variation of the leading exchange couplings. For instance, the interaction path $J$ between two consecutive [CuX$_4$]-tetrahedra in the $b$-direction is controlled by the Cu-X3-X3 angle $\alpha $, the Cu-X3 distance and the Cu-X3-X3-Cu path length (see Table~\ref{tab3}). Remarkably, the variation  with temperature of these particular structural parameters is significantly pronounced for Cs$_2$CuCl$_{4}$ and most probably the reason for the strong temperature dependence of $J$ and $J'$ evident from Fig.~\ref{fig6}. 

\section{Conclusions}
We have presented a detailed low-temperature structural characterization of the triangular antiferromagnets Cs$_2$Cu(Cl$_{4-x}$Br$_{x}$). Our measurements show a site-selective occupation of  Br atoms on specific halogen sites, leading to a strong variation of the local Cu environment as a function of temperature and Br-concentration. The temperature-dependent experimental structure data was then used as a starting point for calculating the magnetic coupling constants using density functional methods. This joint experimental and theoretical study reveals  a strong temperature dependence of the magnetic interactions in Cs$_2$CuCl$_{4}$. Such behavior is caused  by both, a large sensitivity of the exchange interactions to the structural details as well as the particularly non-trivial temperature dependence of the crystal structure. We believe that considering the so far ignored role of temperature-induced variations of spin superexchange couplings in Cs$_2$CuCl$_{4}$ may improve the understanding of its magnetic properties.

\section*{Acknowledgements}
The authors thank  K.-D. Luther for technical assistance. We acknowledge discussions with A.-A. Haghighirad and K. Hradil. This work was supported by the Deutsche Forschungsgemeinschaft through SFB/TRR49. 

\bibliographystyle{apsrev}

\begin{thebibliography}{19}
\expandafter\ifx\csname natexlab\endcsname\relax\def\natexlab#1{#1}\fi
\expandafter\ifx\csname bibnamefont\endcsname\relax
  \def\bibnamefont#1{#1}\fi
\expandafter\ifx\csname bibfnamefont\endcsname\relax
  \def\bibfnamefont#1{#1}\fi
\expandafter\ifx\csname citenamefont\endcsname\relax
  \def\citenamefont#1{#1}\fi
\expandafter\ifx\csname url\endcsname\relax
  \def\url#1{\texttt{#1}}\fi
\expandafter\ifx\csname urlprefix\endcsname\relax\def\urlprefix{URL }\fi
\providecommand{\bibinfo}[2]{#2}
\providecommand{\eprint}[2][]{\url{#2}}

\bibitem[{\citenamefont{Coldea et~al.}(2001)\citenamefont{Coldea, Tennant,
  Tsvelik, and Tylczynski}}]{Coldea:2001}
\bibinfo{author}{\bibfnamefont{R.}~\bibnamefont{Coldea}},
  \bibinfo{author}{\bibfnamefont{D.~A.} \bibnamefont{Tennant}},
  \bibinfo{author}{\bibfnamefont{A.~M.} \bibnamefont{Tsvelik}},
  \bibnamefont{and}
  \bibinfo{author}{\bibfnamefont{Z.}~\bibnamefont{Tylczynski}},
  \bibinfo{journal}{Phys. Rev. Lett.} \textbf{\bibinfo{volume}{86}},
  \bibinfo{pages}{1335} (\bibinfo{year}{2001}).

\bibitem[{\citenamefont{Coldea et~al.}(2002)\citenamefont{Coldea, Tennant,
  Habicht, Smeibidl, Wolters, and Tylczynski}}]{Coldea:2002}
\bibinfo{author}{\bibfnamefont{R.}~\bibnamefont{Coldea}},
  \bibinfo{author}{\bibfnamefont{D.~A.} \bibnamefont{Tennant}},
  \bibinfo{author}{\bibfnamefont{K.}~\bibnamefont{Habicht}},
  \bibinfo{author}{\bibfnamefont{P.}~\bibnamefont{Smeibidl}},
  \bibinfo{author}{\bibfnamefont{C.}~\bibnamefont{Wolters}}, \bibnamefont{and}
  \bibinfo{author}{\bibfnamefont{Z.}~\bibnamefont{Tylczynski}},
  \bibinfo{journal}{Phys. Rev. Lett.} \textbf{\bibinfo{volume}{88}},
  \bibinfo{pages}{137203} (\bibinfo{year}{2002}).

\bibitem[{\citenamefont{Coldea et~al.}(2003)\citenamefont{Coldea, Tennant, and
  Tylczynski}}]{Coldea:2003}
\bibinfo{author}{\bibfnamefont{R.}~\bibnamefont{Coldea}},
  \bibinfo{author}{\bibfnamefont{D.~A.} \bibnamefont{Tennant}},
  \bibnamefont{and}
  \bibinfo{author}{\bibfnamefont{Z.}~\bibnamefont{Tylczynski}},
  \bibinfo{journal}{Phys. Rev. B} \textbf{\bibinfo{volume}{68}},
  \bibinfo{pages}{134424} (\bibinfo{year}{2003}).

\bibitem[{\citenamefont{Radu et~al.}(2005)\citenamefont{Radu, Wilhelm,
  Yushankhai, Kovrizhin, Coldea, Tylczynski, L\"uhmann, and
  Steglich}}]{Radu:2005a}
\bibinfo{author}{\bibfnamefont{T.}~\bibnamefont{Radu}},
  \bibinfo{author}{\bibfnamefont{H.}~\bibnamefont{Wilhelm}},
  \bibinfo{author}{\bibfnamefont{V.}~\bibnamefont{Yushankhai}},
  \bibinfo{author}{\bibfnamefont{D.}~\bibnamefont{Kovrizhin}},
  \bibinfo{author}{\bibfnamefont{R.}~\bibnamefont{Coldea}},
  \bibinfo{author}{\bibfnamefont{Z.}~\bibnamefont{Tylczynski}},
  \bibinfo{author}{\bibfnamefont{T.}~\bibnamefont{L\"uhmann}},
  \bibnamefont{and} \bibinfo{author}{\bibfnamefont{F.}~\bibnamefont{Steglich}},
  \bibinfo{journal}{Phys. Rev. Lett.} \textbf{\bibinfo{volume}{95}},
  \bibinfo{pages}{127202} (\bibinfo{year}{2005}).

\bibitem[{\citenamefont{Ono et~al.}(2003)\citenamefont{Ono, Tanaka,
  Aruga~Katori, Ishikawa, Mitamura, and Goto}}]{Ono:2003}
\bibinfo{author}{\bibfnamefont{T.}~\bibnamefont{Ono}},
  \bibinfo{author}{\bibfnamefont{H.}~\bibnamefont{Tanaka}},
  \bibinfo{author}{\bibfnamefont{H.}~\bibnamefont{Aruga~Katori}},
  \bibinfo{author}{\bibfnamefont{F.}~\bibnamefont{Ishikawa}},
  \bibinfo{author}{\bibfnamefont{H.}~\bibnamefont{Mitamura}}, \bibnamefont{and}
  \bibinfo{author}{\bibfnamefont{T.}~\bibnamefont{Goto}},
  \bibinfo{journal}{Phys. Rev. B} \textbf{\bibinfo{volume}{67}},
  \bibinfo{pages}{104431} (\bibinfo{year}{2003}).

\bibitem[{\citenamefont{Ono et~al.}(2005)\citenamefont{Ono, Tanaka, Nakagomi,
  Kolomiyets, Mitamura, Ishikawa, Goto, Nakajima, Oosawa, Koike
  et~al.}}]{Ono:2005}
\bibinfo{author}{\bibfnamefont{T.}~\bibnamefont{Ono}},
  \bibinfo{author}{\bibfnamefont{H.}~\bibnamefont{Tanaka}},
  \bibinfo{author}{\bibfnamefont{T.}~\bibnamefont{Nakagomi}},
  \bibinfo{author}{\bibfnamefont{O.}~\bibnamefont{Kolomiyets}},
  \bibinfo{author}{\bibfnamefont{H.}~\bibnamefont{Mitamura}},
  \bibinfo{author}{\bibfnamefont{F.}~\bibnamefont{Ishikawa}},
  \bibinfo{author}{\bibfnamefont{T.}~\bibnamefont{Goto}},
  \bibinfo{author}{\bibfnamefont{K.}~\bibnamefont{Nakajima}},
  \bibinfo{author}{\bibfnamefont{A.}~\bibnamefont{Oosawa}},
  \bibinfo{author}{\bibfnamefont{Y.}~\bibnamefont{Koike}},
  \bibnamefont{et~al.}, \bibinfo{journal}{J. Phys. Soc. Jpn.}
  \textbf{\bibinfo{volume}{74}}, \bibinfo{pages}{135} (\bibinfo{year}{2005}).

\bibitem[{\citenamefont{Bailleul et~al.}(1991)\citenamefont{Bailleul, Svoronos,
  Porcher, and Tomas}}]{Bailleul:1991}
\bibinfo{author}{\bibfnamefont{S.}~\bibnamefont{Bailleul}},
  \bibinfo{author}{\bibfnamefont{D.}~\bibnamefont{Svoronos}},
  \bibinfo{author}{\bibfnamefont{P.}~\bibnamefont{Porcher}}, \bibnamefont{and}
  \bibinfo{author}{\bibfnamefont{A.}~\bibnamefont{Tomas}}, \bibinfo{journal}{C.
  R. Hebd. Acad. Sci.} \textbf{\bibinfo{volume}{313}}, \bibinfo{pages}{1149}
  (\bibinfo{year}{1991}).

\bibitem[{\citenamefont{Morosin and C}(1960)}]{Morosin:1960}
\bibinfo{author}{\bibfnamefont{B.}~\bibnamefont{Morosin}} \bibnamefont{and}
  \bibinfo{author}{\bibfnamefont{L.~E.} \bibnamefont{C}},
  \bibinfo{journal}{Acta Crystallogr.} \textbf{\bibinfo{volume}{13}},
  \bibinfo{pages}{807} (\bibinfo{year}{1960}).

\bibitem[{\citenamefont{Kr\"uger et~al.}(2010)\citenamefont{Kr\"uger, Belz,
  Schossau, Haghighirad, Cong, Wolf, Gottlieb-Schoenmeyer, Ritter, and
  Assmus}}]{Kruger:2010}
\bibinfo{author}{\bibfnamefont{N.}~\bibnamefont{Kr\"uger}},
  \bibinfo{author}{\bibfnamefont{S.}~\bibnamefont{Belz}},
  \bibinfo{author}{\bibfnamefont{F.}~\bibnamefont{Schossau}},
  \bibinfo{author}{\bibfnamefont{A.~A.} \bibnamefont{Haghighirad}},
  \bibinfo{author}{\bibfnamefont{P.}~\bibnamefont{Cong}},
  \bibinfo{author}{\bibfnamefont{B.}~\bibnamefont{Wolf}},
  \bibinfo{author}{\bibfnamefont{S.}~\bibnamefont{Gottlieb-Schoenmeyer}},
  \bibinfo{author}{\bibfnamefont{F.}~\bibnamefont{Ritter}}, \bibnamefont{and}
  \bibinfo{author}{\bibfnamefont{W.}~\bibnamefont{Assmus}},
  \bibinfo{journal}{Cryst. Growth Des.} \textbf{\bibinfo{volume}{10}},
  \bibinfo{pages}{4456} (\bibinfo{year}{2010}).

\bibitem[{\citenamefont{Cong et~al.}(2011)\citenamefont{Cong, Wolf, de~Souza,
  Kr\"uger, Haghighirad, Gottlieb-Schoenmeyer, Ritter, Assmus, Opahle,
  Foyevtsova et~al.}}]{Cong:2011}
\bibinfo{author}{\bibfnamefont{P.~T.} \bibnamefont{Cong}},
  \bibinfo{author}{\bibfnamefont{B.}~\bibnamefont{Wolf}},
  \bibinfo{author}{\bibfnamefont{M.}~\bibnamefont{de~Souza}},
  \bibinfo{author}{\bibfnamefont{N.}~\bibnamefont{Kr\"uger}},
  \bibinfo{author}{\bibfnamefont{A.}~\bibnamefont{Haghighirad}},
  \bibinfo{author}{\bibfnamefont{S.}~\bibnamefont{Gottlieb-Schoenmeyer}},
  \bibinfo{author}{\bibfnamefont{F.}~\bibnamefont{Ritter}},
  \bibinfo{author}{\bibfnamefont{W.}~\bibnamefont{Assmus}},
  \bibinfo{author}{\bibfnamefont{I.}~\bibnamefont{Opahle}},
  \bibinfo{author}{\bibfnamefont{K.}~\bibnamefont{Foyevtsova}},
  \bibnamefont{et~al.}, \bibinfo{journal}{Phys. Rev. B}
  \textbf{\bibinfo{volume}{83}}, \bibinfo{pages}{064425}
  (\bibinfo{year}{2011}).

\bibitem[{\citenamefont{Foyevtsova et~al.}(2011)\citenamefont{Foyevtsova,
  Opahle, Zhang, Jeschke, and Valent\'i}}]{Foyevtsova:2011}
\bibinfo{author}{\bibfnamefont{K.}~\bibnamefont{Foyevtsova}},
  \bibinfo{author}{\bibfnamefont{I.}~\bibnamefont{Opahle}},
  \bibinfo{author}{\bibfnamefont{Y.~Z.} \bibnamefont{Zhang}},
  \bibinfo{author}{\bibfnamefont{H.}~\bibnamefont{Jeschke}}, \bibnamefont{and}
  \bibinfo{author}{\bibfnamefont{R.}~\bibnamefont{Valent\'i}},
  \bibinfo{journal}{Phys. Rev. B} \textbf{\bibinfo{volume}{83}},
  \bibinfo{pages}{125126} (\bibinfo{year}{2011}).

\bibitem[{\citenamefont{Larson and von Dreele}(2000)}]{GSAS}
\bibinfo{author}{\bibfnamefont{A.}~\bibnamefont{Larson}} \bibnamefont{and}
  \bibinfo{author}{\bibfnamefont{R.}~\bibnamefont{von Dreele}},
  \bibinfo{journal}{Los Alamos National Laboratory Report}
  \textbf{\bibinfo{volume}{86}}, \bibinfo{pages}{748} (\bibinfo{year}{2000}).

\bibitem[{\citenamefont{Pott and Schefzyk}(1983)}]{Pott:1983}
\bibinfo{author}{\bibfnamefont{R.}~\bibnamefont{Pott}} \bibnamefont{and}
  \bibinfo{author}{\bibfnamefont{R.}~\bibnamefont{Schefzyk}},
  \bibinfo{journal}{J. Phys. E} \textbf{\bibinfo{volume}{16}},
  \bibinfo{pages}{445} (\bibinfo{year}{1983}).

\bibitem[{\citenamefont{Petricek et~al.}(2014)\citenamefont{Petricek, Dusek,
  and Palatinus}}]{Jana2006}
\bibinfo{author}{\bibfnamefont{V.}~\bibnamefont{Petricek}},
  \bibinfo{author}{\bibfnamefont{M.}~\bibnamefont{Dusek}}, \bibnamefont{and}
  \bibinfo{author}{\bibfnamefont{L.}~\bibnamefont{Palatinus}},
  \bibinfo{journal}{Z. Kristallogr.} \textbf{\bibinfo{volume}{229}},
  \bibinfo{pages}{345} (\bibinfo{year}{2014}).

\bibitem[{\citenamefont{van Well}(2014)}]{vanWell:2014}
\bibinfo{author}{\bibfnamefont{N.}~\bibnamefont{van Well}}
  (\bibinfo{year}{2014}), \bibinfo{note}{Dissertation, Goethe-University
  Frankfurt (2014).}

\bibitem[{\citenamefont{Tylczynski et~al.}(1992)\citenamefont{Tylczynski,
  Piskunowicz, Nasyrov, Karaev, Schodiev, and Gulamov}}]{Tylczynski:1992}
\bibinfo{author}{\bibfnamefont{Z.}~\bibnamefont{Tylczynski}},
  \bibinfo{author}{\bibfnamefont{P.}~\bibnamefont{Piskunowicz}},
  \bibinfo{author}{\bibfnamefont{A.~N.} \bibnamefont{Nasyrov}},
  \bibinfo{author}{\bibfnamefont{A.~D.} \bibnamefont{Karaev}},
  \bibinfo{author}{\bibfnamefont{K.~T.} \bibnamefont{Schodiev}},
  \bibnamefont{and} \bibinfo{author}{\bibfnamefont{G.}~\bibnamefont{Gulamov}},
  \bibinfo{journal}{Phys. Stat. Sol. (a)} \textbf{\bibinfo{volume}{33}},
  \bibinfo{pages}{133} (\bibinfo{year}{1992}).

\bibitem[{\citenamefont{Koepernik and Eschrig}(1999)}]{FPLO}
\bibinfo{author}{\bibfnamefont{K.}~\bibnamefont{Koepernik}} \bibnamefont{and}
  \bibinfo{author}{\bibfnamefont{H.}~\bibnamefont{Eschrig}},
  \bibinfo{journal}{Phys. Rev. B} \textbf{\bibinfo{volume}{59}},
  \bibinfo{pages}{1743} (\bibinfo{year}{1999}).

\bibitem[{\citenamefont{Perdew et~al.}(1996)\citenamefont{Perdew, Burke, and
  Enzerhof}}]{GGA}
\bibinfo{author}{\bibfnamefont{J.~P.} \bibnamefont{Perdew}},
  \bibinfo{author}{\bibfnamefont{K.}~\bibnamefont{Burke}}, \bibnamefont{and}
  \bibinfo{author}{\bibfnamefont{M.}~\bibnamefont{Enzerhof}},
  \bibinfo{journal}{Phys. Rev. Lett.} \textbf{\bibinfo{volume}{77}},
  \bibinfo{pages}{3865} (\bibinfo{year}{1996}).

\bibitem[{\citenamefont{Eschrig et~al.}(2003)\citenamefont{Eschrig, Koepernik,
  and Chaplygin}}]{GGA+U}
\bibinfo{author}{\bibfnamefont{H.}~\bibnamefont{Eschrig}},
  \bibinfo{author}{\bibfnamefont{K.}~\bibnamefont{Koepernik}},
  \bibnamefont{and}
  \bibinfo{author}{\bibfnamefont{I.}~\bibnamefont{Chaplygin}},
  \bibinfo{journal}{J. Solid State Chem.} \textbf{\bibinfo{volume}{176}},
  \bibinfo{pages}{482} (\bibinfo{year}{2003}).

\end{thebibliography}

\end{document}